# Far-field three-dimensional deep-subwavelength focal spot with azimuthal polarization

Zhongquan Nie[1,*], Jiawei Liu[1], Xiaofei Liu[2], Weichao Yan[2], Yanxiang Zhang[1], Yanting Tian[1], Shaoding Liu[1], Baohua Jia[3,*]

[1]Key Lab of Advanced Transducers and Intelligent Control System, Ministry of Education and Shanxi Province, College of Physics and Optoelectronics, Taiyuan University of Technology, Taiyuan 030024, China

[2]Department of Physics, Harbin Institute of Technology, Harbin 150001, PR China

[3]Centre for Micro-Photonics, Faculty of Science, Engineering and Technology, Swinburne University of Technology, Hawthorn, Victoria 3122, Australia

E-mail address: niezhongquan1018@163.com; bjia@swin.edu.au

## Abstract

This work focuses on the generation of far-field super-resolved pure-azimuthal focal field based on the fast Fourier transform. A self-designed differential filter is first pioneered to robustly reconfigure a doughnut-shaped azimuthal focal field into a bright one with a sub-wavelength lateral scale (~$0.392\lambda$), which offers a 27.3% reduction ratio relative to that of tightly focused azimuthal polarization modulated by a spiral phase plate. By further uniting the versatile differential filter with spatially shifted beam approach, in addition to allowing for an extremely sharper focal spot, whose size is in turn reduced to $0.228\lambda$ and $0.286\lambda$ in the transverse as well as axial directions, the parasitic sidelobes are also lowered to an inessential level (< 20%), thereby enabling an excellent three-dimensional deep-subwavelength focal field (~$\lambda^3/128$). The relevant phase profiles are further exhibited to unravel the annihilation of field singularity and locally linear (i.e. azimuthal) polarization. Our scheme opens a promising route toward efficiently steer and tailor the redistribution of the focal field.

## 1. Introduction

Faced with the ever-growing importance of integrated nano-photonic devices, it is imperative to focus light beams into the smallest possible focal volume. However, the electromagnetic wave nature of light generally leads to the spatial resolution below the order of half wavelength. Numerous strategies [1-5] have been initiated to alleviate this issue, known as the diffraction limit, thus far. Among these approaches, tight-focusing cylindrical vector beams have been proved to be powerful in not only triggering super-resolution focal fields, but also shaping prescribed polarization profiles, owing to its unique vectorial feature originated from the cylindrical symmetry in polarization [6-8]. It is shown that, for the radially polarized beam focused by a high numerical aperture (NA) objective, an appreciably longitudinal polarized field with sub-wavelength lateral scale in the focal region can be formed by means of either binary phase encoding or amplitude modulation [9, 10]. More impressively, the radial incident polarization in a typical $4\pi$ high NA focusing configuration is capable of yielding a three-dimensional (3D) super-resolved longitudinal focal field consisting in focusing light in a coherent fashion through two objective lenses facing each other [11-13]. These versatile longitudinal focal patterns can find broad applications in particle manipulation [14, 15], optical data storage [16, 17], high-resolution imaging [18], lithography [19], and fabrication of 3D photonics crystals [20], among other feats.

On the other hand, a pure-azimuthal bright field that can dictate tangential force is highly desired for rotation of trapped particles in optical manipulation domain [21-23]. However, the azimuthally polarized (AP) field normally presents a wide doughnut-shaped focal field with purely azimuthal polarization due to the destructive interference between its orthogonal transverse components [24, 25]. To cope with this dilemma, the routine spiral phase plate (SPP) is subtly exploited to tailor the AP radiation that permits one to garner the super-resolved pure-transverse (radial + azimuthal) focal fields [26-31], which are favorable for an extensive wealth of potential applications. In addition, researches exhibit the capability of giving rise to a spherical pure-transverse spot with equal 3D spatial resolution through highly focusing the AP vortex beams in the $4\pi$ optical microscopic setup [32]. Despite these intriguing properties, however, the resultant transverse focal fields suffer from either far from perfect focusing resolution (>0.4$\lambda$) [27, 28, 30] or relatively severe sidelobes (> 30%)[26-31] due to the existence of the parasitic radial component along with the intrinsic modulation limitation of the SPP. In this context, it is highly desired to develop novel recipes for eliminating the radial field to achieve purely azimuthal polarization while guaranteeing a trivial side-lobe level (< 20%) by tightly focusing the judiciously tailored AP light fields. More meaningfully, for some practical applications, especially for ultra-high density data storage, configuring a light field to ultra-compact 3D volume far beyond the diffraction limit (typically < 0.3$\lambda$) is of crucial importance.

In this work, we demonstrate a simple yet powerful roadmap that aims at generating the super-resolved pure-azimuthal bright field in the focal region by replacing the SPP with a well-designed differential filter (DF) under the illumination of the AP Bessel-Gaussian (BG) beam [27] in a high NA focusing geometry. Based on this ingenious paradigm, the spatially shifted beam approach is adopted to further ameliorate the focusing performances. It is revealed that the AP focal spot size in both the transverse and axial directions is suppressed to a deep-subwavelength scale (< 0.29$\lambda$), as well as the sidelobes is adjusted to an insignificant level (< 20%), thus giving rise to a prominent 3D deep-subwavelength pure-azimuthal focal field.

## 2. Focal light field modulated by DF

Figure 1 conceptually depicts the schematic configuration of the proposed set-up to yield a super-resolved AP bright field (see the right inset) in the focal region. An incoming linearly polarized (LP) BG beam firstly passes through a collimate system and is subsequently modulated by the spatial light modulator [33], where a self-designed DF can be encoded flexibly. The desired DF is represented as the difference between two cosine functions with a subtle frequency shift (see Figs. 2(a) and (b)). Mathematically, the phase formula of the proposed DF is given by $f(r_0) = cos2\pi(f_0 +\varepsilon)r_0 - cos2\pi f_0 r_0$, where $f_0$ is an initial frequency, as well as $\varepsilon$ stands for an infinitesimal factor relative to $f_0$. After this, the tailored LP beam can be converted into the AP donut-shaped one by virtue of the vogue polarization converter [4, 5], which is further gathered by a high NA objective lens. In principle, the focal field of the phase-modulated AP beam can be numerically evaluated by the vectorial Debye diffraction theory [8, 9, 34, 35]. To illustrate the superiority of the DF for achieving a bright AP focal field, the canonical vector diffraction integral should be converted into the fast Fourier transform of the weight field $\mathbf{E}_t(\theta, \varphi)$ with a longitudinal angular spectrum component [36]. Therefore, the resultant electric field in the focal region can be derived as,

$$\begin{aligned}\mathbf{E}(x,y,z) &= \iint \left[ f(r_0)\mathbf{E}_t(\theta,\varphi)\cos(ik_z z)/\cos\theta \right] \exp\left[-i(k_x x + k_y y)\right] dk_x dk_y \\ &= \mathrm{FT}\{f(r_0)\} * \mathrm{FT}\{\mathbf{E}_t(\theta,\varphi)\cos(ik_z z)/\cos\theta\},\end{aligned} \quad (1)$$

where $r_0$ is the transverse scale in the pupil plane, $\theta$ is the half convergence angle of the objective lens, $\varphi$ is the azimuthal angle on the transverse plane, FT{·} denotes the Fourier transform, the symbol $*$ represents the convolution operator, and $\mathbf{k} = (k_x, k_y, k_z)$ is the wave vector. The weight field $\mathbf{E}_t(\theta, \varphi)$ is the transmitted field after the objective lens, which is given by,

$$\mathbf{E}_t(\theta,\varphi) = \begin{bmatrix} E_x \\ E_y \end{bmatrix} = \cos^{1/2}\theta \exp\left[-\left(\frac{\sin\theta}{\sin\alpha}\right)^2\right] J_1\left(\frac{2\sin\theta}{\sin\alpha}\right) \begin{bmatrix} -\sin\varphi \\ \cos\varphi \end{bmatrix}, \quad (2)$$

where $\alpha$ denotes the maximum focal angle determined by the NA of the objective lens, and $J_1$ is the first kind of Bessel function of order 1. From Eq. (1), one can see that the focused field is described as a spatial convolution with the Fourier transform of $f(r_0)$ and the Fourier transform of the unmodulated field related to $\mathbf{E}_t(\theta, \varphi)$ (which is a hollow AP field governed by $J_1(\theta)$). Taking into account both the recursion formula of the Bessel function ($2J_1'(t) = J_0(t) - J_2(t)$, in which $J_1'(t)$ indicates the derivative of $J_1(t)$ with respect to $t$) and translation trait of the $\delta$ function ($G(t)*\delta(t-t_0) = G(t-t_0)$, where $G(t)$ is an arbitrary function, and $t_0$ is the translation space), it is therefore seems to be of viable track to allow us to garner the bright AP focal field provided that the Fourier transform of $f(r_0)$ reveals the difference between two $\delta$ functions with tiny shift in the spectrum plane [37]. Executing some straightforward operations, the first term in Eq. (1) can be expressed as,

$$\text{FT}\{f(r_0)\} = \frac{1}{2}\{\delta[r+(f_0+\varepsilon)]-\delta(r+f_0)\} + \frac{1}{2}\{\delta[r-(f_0+\varepsilon)]-\delta(r-f_0)\}, \quad (3)$$

where $r = (x^2+y^2)^{1/2}$ stands for the transverse scale in the focal region. As expected, the Fourier transform of $f(r_0)$ consists of the superposition of two comparable terms associated with the differences between the prescribed $\delta$ functions (see Fig. 2(c)). From the above analysis we can realize a solid AP spot with sub-wavelength scale instead of opponent dark one by integrating Eq. (3) with Eq. (1), corresponding to the differential of donut-shaped focal field without undergoing the versatile DF along a symmetry direction $r$. In our calculations, the parameters are selected as: the wavelength $\lambda = 633$ nm, NA = 0.95, the refractive index $n_0 = 1$, the initial frequency $f_0 = 100$/mm, $\varepsilon = 1$/mm, and the half convergence angle $0 \leq \theta \leq \sin^{-1}(\text{NA}/n_0)$.

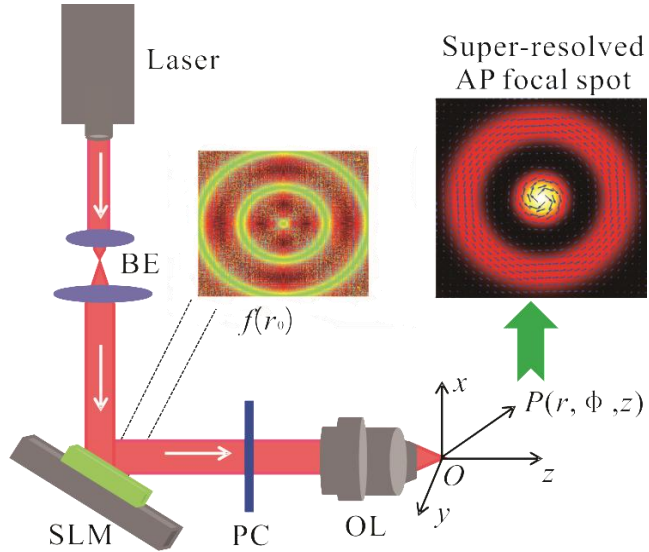

Fig. 1. Schematic of the set-up to create super-resolved AP field. Left insets represent the phase profile of the DF in the $x$-$y$ plane. Right inset exhibits the generated super-resolved AP focal spot at $z = 0$, in which the blue arrows indicate the polarization sate distribution. BE: beam expander, SLM: spatial light modulator, PC: polarization converter, OL: objective lens.

As predicted above, the proposed DF plays a crucial role in reconfiguring the focusing pattern of the AP BG beam. To explore the beam-shaping performance of this versatile DF, numerical calculations are performed in accordance with Eqs. (1)-(3). Figures 2(d) and 2(e) give, respectively, the electrical intensity profiles in the $r$-$z$ plane and in the focal region for the focusing system with the DF. It is shown that, under such a circumstance, a bright focal spot is created expectedly and the polarization direction is fully tangential (blue arrows in Fig. 2(e)), indicating that a completely AP bright field is hands-down. This scenario is entirely alien from the work in [24, 25], which features a doughnut-shaped intensity patterns with purely azimuthal polarization due to the existence of polarization singularity. Besides, it is interesting to note that the polarization orientation of the inner spot is clockwise whereas the polarization orientation of the outer ring becomes anticlockwise as shown in Fig. 2(c), which is propitious to synchronously rotate and revolve multiple particles for a fixed distance to the focus along opposite tangential directions [21-23].

To offer more insight into the focusing ability of the AP BG beam modulated by the assigned DF, the normalized intensity distribution in the transverse direction is plotted in Fig. 2(f) (red curve), from which the full width at half-maximum (FWHM) value in the lateral direction ($W_r$) is determined to be $0.392\lambda$. Apparently, such a sharp value enormously overwhelms the intrinsic diffraction limit $\lambda/2NA=0.526\lambda$, thus giving birth to a super-resolved pure-azimuthal focal field. In striking contrast, the $W_r$ for the highly focused AP BG beam with undergoing SPP extends up to $0.539\lambda$ (black curve), which is broadened by 37.5% in comparison with that of the DF modulation. The underlying reason for this difference is thought to be responsible for the emergence of additional radial field component for SPP modulated AP beam that will broaden the total spot size for sure [26-28]. Even more, our proposed scheme is always superior to those focusing situations of radially polarized beam in terms of the focal spot scale [4, 8, 9]. Physically, the essential reason to produce the lateral super-resolved focal field with purely azimuthal polarization is that the versatile DF, unlike the SPP, only serves as a special amplitude-reconstruction filter but is polarization-independent, which gathers the light away from the center of radiation field and diverts a whole host of energies towards the vicinity of the focus, thus making the bright super-resolved AP field possible.

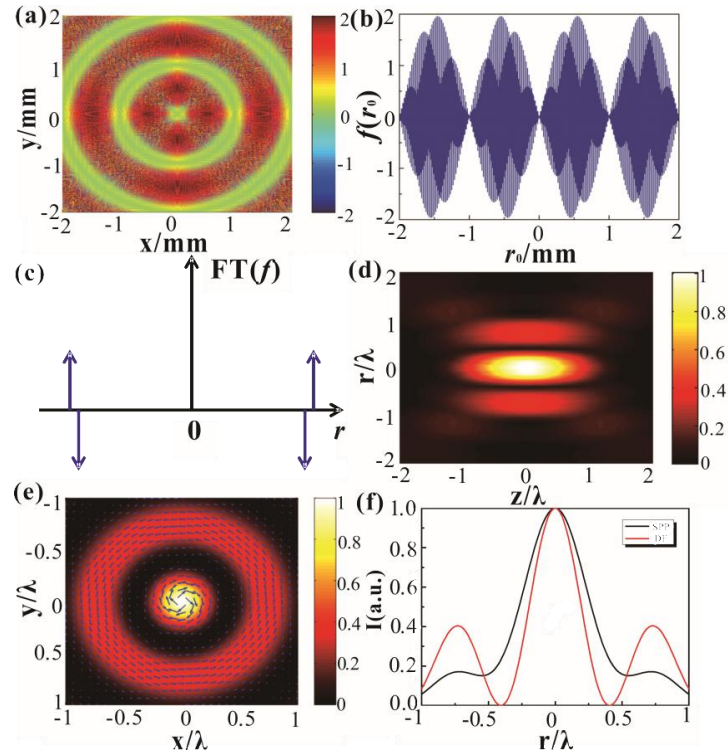

Fig. 2. Phase distribution of the DF and the focusing profile of the DF modulated AP BG beam. (a) The phase pattern $f(r_0)$ in the x-y plane; (b) $f(r_0)$ as a function of $r_0$; (c) Fourier transform of $f(r_0)$; (d) and (e) contour plots of the intensity distributions in the r-z and x-y plane after the DF

modulation; (f) the normalized electric intensity with SPP (black curve) and DF (red curve) modulation.

## 3. Alliance between DF and shift beam approach tailor focal field

Apart from the central focal spot, appreciable transverse sidelobes ($S_r$) emerge with an intensity of around 40.4% of the principle lobe peak value, as revealed in 2(f) (red curve). Such redundant sidelobes, in practice, degenerate the quality of focused DF-loaded AP beam and bring about a great background noise, which does not favor high-density optical data storage, high-resolution imaging, and lithography. It is therefore highly desired to develop novel strategies to deal with the undesirable sidelobes. Numerous efforts have been made to achieve perfect focusing with suppressed sidelobes until now, for example, using Toraldo filters [38], shifted beam approach [39], and metamaterial lens [40]. Among these approaches, using spatially shifted beam is the preferential route, owing to the capability to lower the sidelobes and simultaneously sharpen the focal spot.

As a beneficial extension of the work in [39], on the one hand, the host fields in the peripheral regions overlap in proportion to the shifted beams from the central regions, which is conductive to lowering the undesired side-lobe levels. On the other hand, the side-lobe field of the shifted beams can be designed to destructively interfere with the main-lobe profile of the original field and yield a null in the total field close to the peak of the principle lobe, leading to a sharper focal spot dimension. In these regards, integrating the DF modulation with the shifted beam approach, we are able to garner 3D super-resolved AP focal field with suppressed side-lobe levels in principle.

To achieve this, an AP light source as a whole should be shifted along both the $r$ and $z$ directions, respectively. As a demonstration, we assume that the initial radiation field is, respectively, migrated from the origin to two new positons $r = r_+$ and $r = r_-$ along the positive and negative $r$ axis. Likewise, the same incident field, in turn, moves to $z = z_+$ and $z = z_-$ along the optical axis. Under both the original and shifted AP beams incidence, the electric field pattern near focus produced by single light source can be extended to a superposition of three focal fields through the fast Fourier transform. As a consequence, the resultant focal field in Eq. (1) should be rewritten as

$$\mathbf{E}_s(x,y,z) = \mathbf{E}(x,y,z) \times \left(1 + \tau_+ \exp\left[i2\pi(\mathbf{r}-r_+)/\lambda\right] + \tau_- \exp\left[i2\pi(\mathbf{r}+r_-)/\lambda\right]\right) \\ \times \left(1 + \delta_+ \exp\left[i2\pi(\mathbf{z}-z_+)/\lambda\right] + \delta_- \exp\left[i2\pi(\mathbf{z}+z_-)/\lambda\right]\right). \quad (4)$$

In Eq. (4), $\tau_\pm$ and $\delta_\pm$ represent, respectively, the tunable weights assigned to individual shifted beam in the transverse and axial directions, which are chosen to rescale the shifted electric field components. The vector manifestations of $\mathbf{r}$ and $\mathbf{z}$ indicate that the migration of the initial light field is dictated by orientations. It should be pointed out that $\mathbf{E}_s(x, y, z)$ is composed of the superposition of three shifted beams in a spatially coherent manner relative to $\mathbf{E}(x, y, z)$. Also, the transverse/longitudinal FWHM ($W_r$/$W_z$) and relevant side-lobe value ($S_r$/$S_z$) of the focal field are merely associated with both the transverse/longitudinal weight ($\tau_\pm/\delta_\pm$) and migrated distance ($r_\pm/z_\pm$). By judiciously adjusting the crucial parameters $r_\pm$, $z_\pm$, $\tau_\pm$ and $\delta_\pm$, perfectly destructive interference thus will be caused by the presence of a $\pi$ phase difference between the original and rescaled shifted fields [41].

Figure 3 shows the $W_r$ ($W_z$) and $S_r$ ($S_z$) of the central peak as a function of $|r_\pm|$ ($|z_\pm|$) for the case of $\tau_\pm=-3$, and $\delta_\pm=-1.5$. It can be seen from Fig. 3(a) that both the $W_r$ and $S_r$ change drastically with the $|r_\pm|$, as well as present inverse trends entirely. Namely, the smaller the $W_r$ is, the larger the $S_r$ is. The maximal $W_r$ (~0.228$\lambda$) and minimal $S_r$ (~13.7%) over the range of $|r_\pm|<\lambda$ emerge at $|r_\pm|=0.5\lambda$. As such, the variation of the $W_z$ and $S_z$ with the $|z_\pm|$ is plotted in Fig. 3 (b). It is revealed that the $S_z$ decreases linearly with increments of the $|z_\pm|$ and then begins to raise symmetrically, whereas the $W_z$ (~0.286$\lambda$) is independent wholly of the $|z_\pm|$. We note that, in this case, the lowest $S_z$ (~18.4%) occurs at $|z_\pm|=0.5\lambda$. In these contexts, we are thus capable of implementing ultra-small-scale (< 0.3$\lambda$) focal fields companying trivial sidelobes (< 20%) using a compatible parameter set.

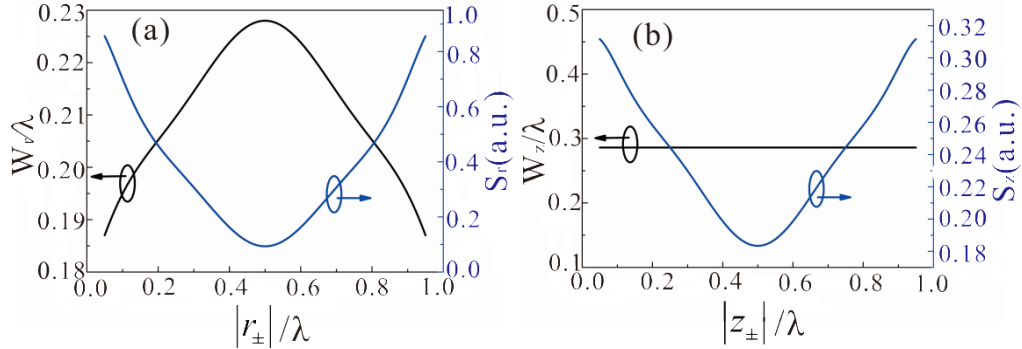

Fig. 3. Dependence of both the focal spot size and side-lobe value on the shift distance. (a) $W_r$ and $S_r$ versus $|r_\pm|$ for the case of $|z_\pm|=0.5\lambda$, and (b) $W_z$ and $S_z$ versus $|z_\pm|$ for the case of $|r_\pm|=0.5\lambda$. The other parameters are $\tau_\pm=-3$, and $\delta_\pm=-1.5$.

Figure 4 illustrates the intensity profiles in the focusing configure integrating the DF modulation with the shifted beam method. In addition to the aforementioned parameters, we perform the integration numerically using the parameters $r_+ = z_+ = 0.5\lambda$, $r_- = z_- = -0.5\lambda$, $\tau_\pm=-3$, and $\delta_\pm=-1.5$ as stated in Fig. 3. As expected, it is observed from Fig. 4(a) that the focal spot size is enormously squeezed in the both $r$ and $z$ directions compared with the case of focal field only related to the DF modulation (see Fig. 2(d)). Furthermore, equally important is that the polarization orientation is fully azimuthal in Fig. 4(b) (see blue arrows). More precisely, the total intensity profile with undergoing the shifted beams in the focal cross-section is shown in Fig. 4(c) (red curve). Both the $W_r$ and $S_r$ are 0.228$\lambda$ and 13.7%, which is, respectively, reduced by 41.8% and 66.1% with respect to focal field with the DF modulation (see black curve in Fig. 4(c)). Analogously, the intensity profile by means of the shifted beam approach along the optical axis is revealed in Fig. 4(d) (red curve), from which both the $W_z$ and $S_z$ are, in turn, evaluated to be $0.286\lambda$ and $18.4\%$. In addition, its axial dimension is 4.5 narrower than that obtained with only DF modulation (~$1.56\lambda$, see black curve in Fig. 4(d)). This scenario is partly different from the classic $4\pi$ optical system [11-13, 42, 43], as our setup doesn't require complex configuration. Therefore, the proposed simple yet powerful roadmap in this work can not merely achieve far-field 3D deep-subwavelength ($< 0.29\lambda$) AP focal field, but also push the sidelobes to a trivial level ($< 20\%$).

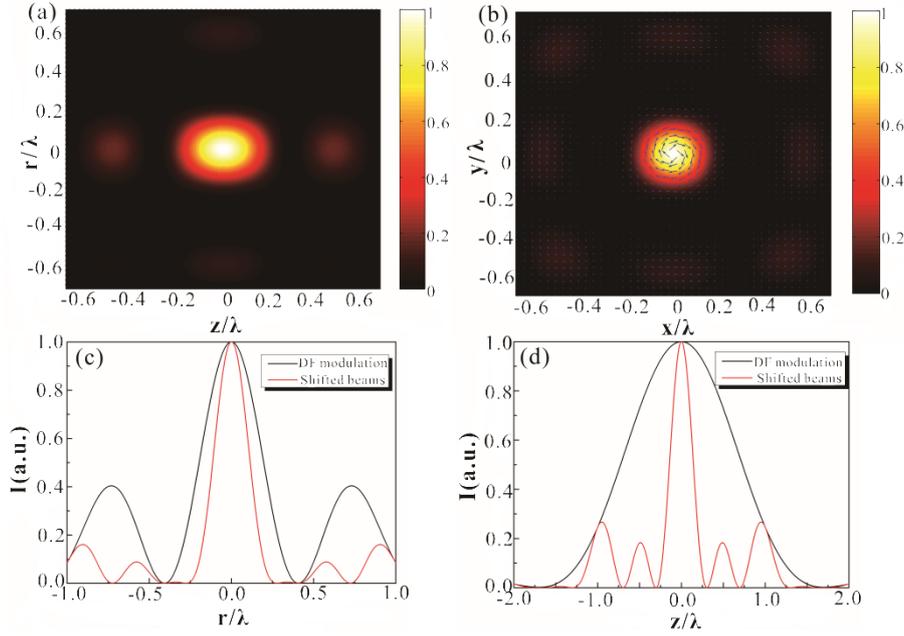

Fig. 4. Intensity distributions of purely AP focal spot from DF modulation with the superposition of shifted beam approach. (a) in the $r$-$z$ plane, and (b) in the focal region ($z = 0$); (c) and (d) the normalized intensity distributions along lateral and axial directions.

To further verify the implementation of bright AP focal field in our proposed scheme, the corresponding phase patterns ($\Phi$) for several modulation scenarios in the $r$-$z$ plane are depicted in Fig. 5. It is noteworthy that, for a light beam with space-variant polarization, the phase distribution can be determined with the aid of the canonical definition by Pancharatnam: $\Phi = \langle \mathbf{E}(\mathbf{r}_1), \mathbf{E}(\mathbf{r}_2) \rangle$, in which $\langle \rangle$ represents an inner product of two electric fields $\mathbf{E}(\mathbf{r}_1)$ and $\mathbf{E}(\mathbf{r}_2)$ at different spatial positions on the wavefront. As a contrast, Fig. 5(a) shows the phase landscape of strongly focused AP beams without any modulation, which exhibits an abrupt change from $\pi$ to 0 at $r = 0$ (see black solid line in Fig. 5(f)), thus giving rise to phase singularity at the beam center. If the AP beams impinging on a first-order vortex phase element are gathered by a high NA objective lens, the phase singularity fully disappears in the focal plane, as displayed in Fig. 5(b). In particular, the phase always remains a constant value of almost $-\pi/2$ within the central regime ($-0.443\lambda<r<0.443\lambda$), whereas it rapidly leaps into another constant of nearly $\pi/2$ on the periphery (see red dash line in Fig. 5(f)). This means that it is able to yield a small-scale quasi-circular polarized beam in the focal plane under this circumstance [44, 45].

In a similar fashion, an intriguing situation can be encountered when our presented DF instead of the SPP revealed in Fig. 5(b) is superimposed on the initial AP beam. From Fig. 5(c), it can be seen that, as expected, the phase profile of the AP beam after DF modulation is energetically immune to singularity it the focal plane. From closer inspection of this undistorted phase distribution, we recognize that it consists of several concentric ring and a phase difference of $\pi$ appears between the adjacent rings. Another interesting observation is that the intrinsic phase holds a constant of 0 all the time within a lateral realm ranging from $-0.404\lambda$ to $0.404\lambda$ (see blue dot line in Fig. 5(f)), manifesting that a sub-wavelength bright field with locally linear (i.e. azimuthal) polarization is allowable. What's more, the undistorted phase pattern is further suppressed to a deep-subwavelength scope of $|r|<0.273\lambda$ (see magenta dash dot line in Fig. 5(f)) by the alliance between the SPP (see Fig. 5(c)) and the shifted beam phase (see Fig. 5(d)), as demonstrated in Fig. 5(e). This can be attributed to the fact that the side-lobe region of the shifted beams and main-lobe one of the original field are in reverse phase. Overall, the demonstrated finding implies that the focal pattern and polarization state of the DF-assisted AP beam are, separately, indeed converted to bright field and locally linear (i.e. azimuthal) polarization in the focal region.

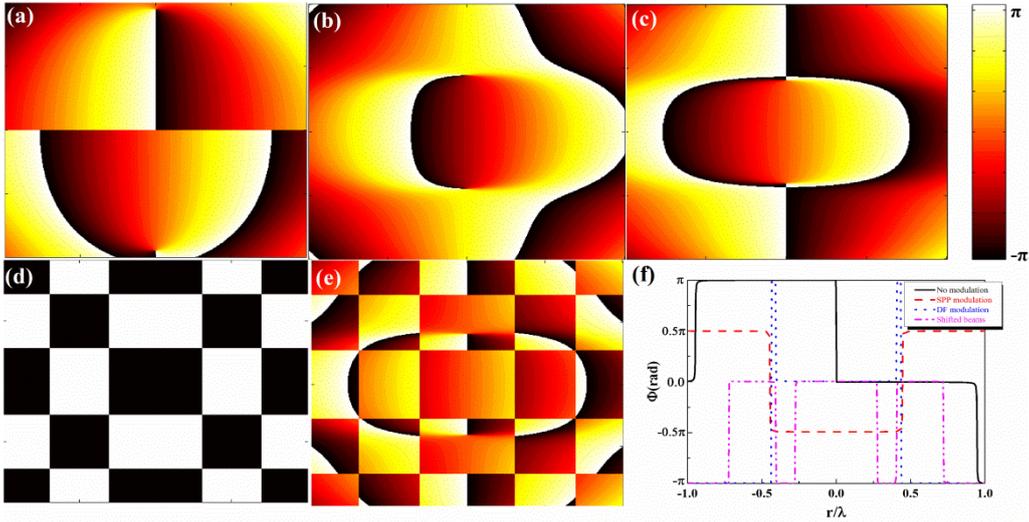

Fig. 5. Phase patterns in the *r-z* plane for tightly focusing AP beams for several modulation scenarios. (a) without any modulation; (b) with SPP modulation; (c) with DF modulation; (d) shifted beam phase distribution; (e) with DF and shifted beam phase modulation; (f) the relevant phase profiles along the lateral direction. The dimension is $1\lambda\times1\lambda$.

The corresponding 3D iso-intensity surfaces of the modulated focal field are shown in Fig. 6. It is remarkable from Fig. 6(a) that the main lobe of the focal AP field modulated by DF is totally encircled by the lateral sidelobes for the case $I(x, y, z) = e^{-1}I_{max}$, in which $I(x, y, z)$ denotes the focal field intensity and $I_{max}$ is the relevant peak intensity. Yet, only a substantially depressed focal spot along with fully disappeared side-lobe level arises, provided that the AP field is subjected to both the DF action and shifted beam scheme simultaneously (see Fig. 6(b)). This is due to the fact that the appointed iso-intensity value ($e^{-1}$) is slightly smaller than the sidelobes of 40.4% in Fig. 2(d) whereas an opposite tendency occurs in Fig. 4(a). In a different front, when we choose $I(x, y, z) = 0.5I_{max}$, the parasitic sidelobes completely vanish in aforesaid either situation (see Figs. 6(c) and 6(d)). To assess the best reachable 3D focal spot resolution in the case of the DF modulation aligning with the shifted beam approach, the voxel size of the focal spot can be calculated as $4\pi/3\times(W_r/2)^2\times W_z/2 = \lambda^3/128$, far inferior than the diffraction limit of $\lambda^3/8$. Such a 3D deep-subwavelength ellipsoid-shaped focal spot is even parallel to that obtained in stimulated emission depletion technique [46-48]. What is really needed to demonstrate that regulating the $|r_\pm|$ can serve to further compress the lateral scale but result in incremental side-lobe levels and a higher energy depletion in the main lobe region (see Fig. 3(a)). Meanwhile, increasing the number of shifted beams can effectively diminish the sidelobes [39]. Therefore, a tradeoff should be involved among focal spot extension, side-lobe level and energy transition efficiency by pondering over diverse shifted separations and the number of shifted beams.

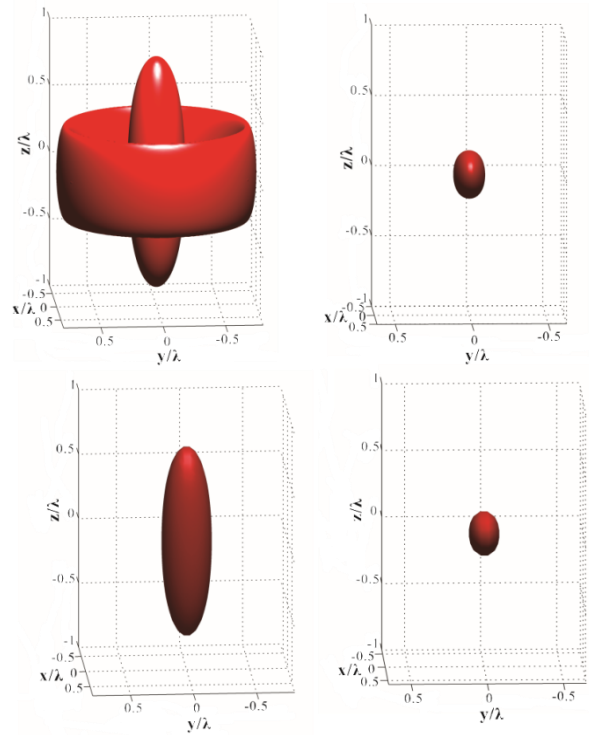

Fig. 6. 3D iso-intensity surfaces given by the intensity at 1/e and the half of the peak value. (a) and (c) with the DF modulation; (b) and (d) with DF modulation and shifted beam approach.

As for the experimental realization of the presented trick, the crucial issues are twofold: (i) the generation of the vector BG beams, and (ii) encoding the desired phase patterns. The former is achieved by transmitting a linearly polarized Gaussian beam through a consecutive assembly of a spatial light modulator, a spiral phase plate, and a polarization converter [49]. Moreover, in order to guarantee exact implementation of the predetermined DF and shifted beam-assisted element, which will perform ideal reconfiguration of the focal field, the experimental available spatial light modulator (for example, Holoeye Pluto) should be at least able to produce 256 levels of synergetic phase modulation in the range of 0-2$\pi$. In response to these, We are currently working on providing experimental demonstrations.

## 4. Conclusion

To conclude, we have firstly presented the generation of a pure-azimuthal focal field with a sub-wavelength lateral scale (~0.392$\lambda$) through tightly focusing an AP BG beam encoded by a well-designed DF. Owing to the superior capability of this binary phase element that behaves as a dedicated amplitude-reconstruction filter, a central bright AP focal field is capable of being successfully accessible couple with relatively substantial sidelobe of 40.4%. More impressively, by further integrating the DF modulation with the spatially shifted beam scheme, we can not only render the focal spot size to be in turn suppressed to 0.228$\lambda$ and 0.286$\lambda$ along both the transverse and axial directions, but also push the sidelobes to a trivial level (< 20%), thus enabling a perfect 3D deep-subwavelength (~$\lambda^3$/128) focal field. The associated phase value of zero in the focal region further confirms the locally linear (i.e. azimuthal) polarization of the solid focal field. In contrast to other alternative technologies regarding focusing properties, for instance, 4$\pi$ optical microscopic setup [11-13, 42, 43, 50], stimulated emission depletion technique [46-48], and absorbance modulation [51, 52], our proposed scheme is equally powerful but simpler circumventing serious practical limitations. The 3D ultra-compact AP focal field is extremely encouraging for the rotation and acceleration of multiple particles, high-density optical data storage, and optical materials and instruments responsive to the azimuthal field only.

## Acknowledgement

This work was supported by the National Natural Science Foundation of China (Nos. 61575139, 11474077, 11374079, 11604236, 51602213, 61605136 and 11404283).